\documentclass[12pt]{article}
\usepackage[latin1]{inputenc}
\usepackage{amsfonts}
\usepackage{amssymb,amsmath}

\def\ep{\varepsilon}

\def\Tr{\mathrm{Tr}}
\def\diag{\mathrm{diag}}
\def\p{\partial}
\def\CC{\mathrm{C}}
\def\TT{\mathrm{T}}
\def\da{{\dot a}}
\def\db{{\dot b}}
\def\ep{\varepsilon}
\def\mn{{\mu\nu}}

\def\({\left\{}
\def\){\right\}}
\def\[{\left[}
\def\]{\right]}
\def\Pc{\mathbb{P}_\mathrm{C}}
\def\Pa{\mathbb{P}_\mathrm{A}}
\def\Pt{{\mathbb{P}_\mathrm{T}}}
\def\Pcm{\mathbb{P}_\mathrm{C}^m}
\def\Pam{\mathbb{P}_\mathrm{A}^m}
\def\Ptm{{\mathbb{P}_\mathrm{T}^m}}
\def\Proy{\mathbb{P}}

\newcommand{\sg}[2]{{{\sigma^{#1}}_{#2}}}
\newcommand{\cor}[1]{\left\{#1\right\}}
\newcommand{\dir}[1]{\left\{#1\right\}_{\textrm{D}}}
\newcommand{\con}[1]{\left[#1\right]}
\newcommand{\du}[2]{_{#1}^{\phantom{#1}#2}}
\newcommand{\ud}[2]{^{#1}_{\phantom{#1}#2}}
\newcommand{\udu}[3]{^{#1\phantom{#2}#3}_{\phantom{#1}#2}}
\newcommand{\mat}[1]{\left(\begin{matrix}#1\end{matrix}\right)}
\newcommand{\bra}[1]{\left|#1\right>}
\newcommand{\ket}[1]{\left<#1\right|}

\begin{document}

\begin{titlepage}

\hfill{Preprint {\bf SB/F/05-333}} \hrule \vskip 2.5cm
\centerline{\bf The quantum algebra of superspace} \vskip 2cm

\centerline{N. Hatcher, A. Restuccia and J. Stephany} \vskip 4mm
\begin{description}
\item[]{\it  Universidad Sim\'on Bol\'{\i}var, Departamento de F\'{\i}sica,
Apartado Postal 89000, Caracas 1080-A, Venezuela.}
\item[]{\it \ \ \ e-mail:  nhatcher@fis.usb.ve, arestu@usb.ve,
 stephany@usb.ve}
\end{description}
\vskip 1cm

\begin{abstract}

We present the complete set of $N=1$, $D=4$ quantum algebras
associated to massive superparticles. We obtain the explicit
solution of these algebras realized in terms of unconstrained
operators acting on the Hilbert space of superfields. These
solutions are expressed using the chiral, anti-chiral and
tensorial projectors which define the three irreducible
representations of the supersymmetry on the superfields. In each
case the space-time variables are non-commuting and their
commutators are proportional to the internal angular momentum of
the representation.  The quantum algebra associated to the chiral
or the anti-chiral projector is the one obtained by the
quantization of the Casalbuoni-Brink-Schwarz (superspin 0) massive
superparticle. We present a new superparticle action for the
(superspin 1/2) case and show that their wave functions are the
ones associated to the irreducible tensor multiplet.

\end{abstract}

\vskip 2cm \hrule
\bigskip
\centerline{\bf UNIVERSIDAD SIMON BOLIVAR} \vfill
\end{titlepage}

\section{Introduction and summary}
The superparticles, the point-like objects which move in the Salam
and Strathdee superspace \cite{SalStr} were first described by
Casalbuoni \cite{Cas2} almost 30 years ago, and by Brink and
Schwarz \cite{BrinkS} soon afterwards. They have internal angular
momentum and built in supersymmetry. As a simple model of a
classical particle with internal angular momentum they bear some
resemblance with the relativistic top studied by Hanson and Regge
\cite{HanReg}, but they also have novel interesting features of it
own. The real interest in these models arose with the advent of
the superstring action \cite{GreeSch} and the understanding of the
structure of  the constraints that appear in the canonical
analysis. Siegel \cite{Sie} showed that the massless superparticle
possess an extra local fermionic symmetry now called kappa
symmetry which is also present in the massive case when central
charges are included \cite{AzL1982}. This symmetry develop later
in a guiding principle that was used to select valid Lagrangeans
for supersymmetric systems. The superstring and the supermembrane
actions were discovered imposing this symmetry. Also it was
discovered that by imposing  the kappa symmetry to the action of a
supermembrane in a background field one obtains the whole set of
eleven dimensional supergravity equations for the component
fields.

But the virtues of supersymmetry and kappa symmetry are
accompanied with problems. The canonical analysis of the massless
superparticle shows that the set of first class constraints that
generate kappa symmetry were tied to a set of second class
constraints. With time it became evident that it was not possible
to split this constraints in a covariant and irreducible way. Over
the years many ways to deal with the problem of covariant
quantization of the superparticle and superstring were proposed
with different grades of success. Many of these methods needed to
include a infinite tower of ghosts and the meaning of the
resulting BRST operator became unclear to say the least. Some
other approaches were linked to the harmonic superspace, where the
set of infinite fields are codified in an organized way
\cite{GalIO2001}. Infinite fields appeared also in the
formulations using twistor variables where as an additional
complication several different versions of what a twistor should
be have been proposed. Very interesting geometrical ideas related
to twistor theory were inherent in the work of doubly
supersymmetric particles and strings for which Siegel symmetry has
been understood as a diffeomorphism of the superworld-line.
Nevertheles, although, extensions for D=10 using Lorentz harmonics
\cite{DelGS1992}, superembeddings techniques \cite{Sor2000} and
more recently pure spinors \cite{Berk} have been proposed, the
goal of constructing off-shell D=10 Super Yang Mills has not been
achieved.

A related but quite different class of systems is that of massive
superparticles. When central charges are absent they do not posses
kappa symmetry and first and second class constraints are not
mixed. The solution of the superparticle algebra  of the
observables of the theory has not been discussed in the
literature. In this paper we solve this problem for $D=4$ and
$N=1$ massive superparticles. We obtain the explicit solution of
these algebras realized in terms of unconstrained operators acting
on the Hilbert space of superfields. These solutions are expressed
using the chiral, anti-chiral and tensorial projectors which
define the three irreducible representations of the supersymmetry
on the superfields.

This paper is organized as follows. In section 2 we discuss the
classical dynamics of the massive superparticle which is subject
to second class constraints. We present the  algebra obtained
using Dirac procedure. The resulting algebra of brackets
\cite{Cas2} is not straightforward to implement quantum
mechanically because the space time variables turns out to be
non-commutative, $\con{x^\mu,x^\nu}\neq 0$. This problem may be
circumvented  by finding a reduced set of coordinates that satisfy
canonical commutation relations which, not surprisingly, result to
be the chiral coordinates as suggested by Casalbuoni \cite{Cas2}.
In section 3 we construct the quantum algebra of the superparticle
with operators acting on the Hilbert space of states corresponding
to the irreducible representations of the supersymmetry. This is
done with the aid of the projectors (chiral, anti-chiral and
tensorial) to the three subspaces of the space of superfields
which allow an irreducible representation of the supersymmetry
algebra (associated respectively to the chiral, anti-chiral and
irreducible tensor multiplets) . We then obtain three well defined
covariant operatorial solutions for the superparticle algebra, two
with superspin 0 associated to the chiral or to the anti-chiral
massive superparticle and one with super spin $1/2$ with should be
associated to a different super particle. In each case we obtain
the explicit operatorial expression for the internal angular
momentum. The chiral and anti-chiral massive superparticles are
described by the standard massive superparticle action, while the
super spin $1/2$ superparticle is described by a different action.
In section 4 we present and analyze the action corresponding to
the super spin $1/2$ superparticle. We show that the wave
functions of this new superparticle  are  superfields projected by
the tensorial projector which reduce them to the degrees of
freedom of the irreducible tensor multiplet (of super spin $1/2$).
In section 5 we discuss the applicability of our results in a more
general setup. Finally in section 6 we present our conclusion and
outlook.
\section{Classical mechanics in the superspace}
\label{chir} Let us consider the standard massive superparticle in
$D=4$. The metric signature is $\eta_\mn=\diag\{-1,+1,+1,+1\}$ and
the superspace coordinates are
$(x^\mu,\theta^{ai},\bar\theta\ud{\da}{i})$, where $a=1,2$ is a
spinor index and $i=1,...,N$ is the number of supersymmetric
charges. Naturally $(\theta^{ai})^*=\bar\theta\ud{\da}{i}$. We
choose Dirac matrices to be off-diagonal and given by,
\begin{gather}
\gamma^\mu=\mat{0 & \sigma^\mu\\ \bar \sigma^\mu & 0}
\end{gather}
with $\sigma\ud{\mu}{a\db}$ the Pauli matrices. The action
principle for the superparticle is given by \cite{Cas2}
\begin{gather}
S=\frac{1}{2}\int
d\tau\left(e^{-1}\omega^\mu\omega^\nu\eta_\mn-em^2\right) \ \ ,
\label{SuperAction}
\end{gather}
where  $\omega^\mu=\dot x^\mu-i\dot\theta^{ai}
\sigma\ud{\mu}{a\db}\bar\theta\ud{\db}{i}+i\theta^{ai}\sigma\ud{\mu}{a\db}
\dot{\bar\theta}\ud{\db}{i}$ is defined for convenience.

The generalized momenta are given by
\begin{gather}
\pi_e=0\\
p_\mu=e^{-1}\omega_\mu\\
\pi_{ai}=-ip_\mu \sigma\ud{\mu}{a\db}\bar\theta\ud{\db}{i}\\
\bar\pi\du{\db}{i}=-ip_\mu\bar\theta^{ai}\sigma\ud{\mu}{a\db}\ \ .
\end{gather}
and satisfy the canonical Poisson bracket relations \cite{Cas2}
\begin{gather}
\cor{x^\mu,p_\nu}=\delta\ud{\mu}{\nu}\\
\cor{\theta^{ai},\pi_{bj}}=-\delta\ud{a}{b}\delta\ud{i}{j}\ \ .
\end{gather}
Along the conserved quantities related to Super Poincar{\`e}
invariance  the total angular momentum  is given by
\begin{gather}
J_\mn=L_\mn+S_\mn\\
L_\mn=x_\mu p_\nu-x_\mu p_\nu\\
S_\mn=-\frac{1}{4}\left(\theta^{ai}{\sigma_\mn}\du{a}{b}\pi_{bi}+
\bar\pi\du{\da}{i}{\bar\sigma_\mn}\,\ud{\da}{\db}\bar\theta\ud{\db}{i}\right)\
\label{ClassicalInternal} \ .
\end{gather}
For this system, there is only one first class constraint
$\pi_e=0$ related to the  reparametrization invariance of the
action which implies the first class secondary constraint
$p^2+m^2=0$. There appear also the  constraints,
\begin{gather}
\label{d1}
d_{ai}\equiv \pi_{ai}+ip_\mu \sigma\ud{\mu}{a\db}\bar\theta\ud{\db}{i}=0 \\
\bar d\du{\db}{i}\equiv \bar\pi\du{\db}{i}+ip_\mu \theta^{ai}\sigma\ud{\mu}{a\db}=0\label{d2}\\
\cor{d_{ai},\bar d\du{\db}{j}}=-i\delta\du{i}{j}p_\mu
\sigma\ud{\mu}{a\db}\equiv {C_{ai\db}}^j\ \ .
\end{gather}
which are second class since the matrix ${C_{ai\db}}^j$ is non
singular if $m\neq 0$. The Dirac brackets are given by,
\begin{gather}
\dir{F,G}=\cor{F,G}-\cor{F,d_{ai}}\hat C\ud{ai\db}{j}\cor{\bar
d\du{\db}{j},G}-\cor{F,\bar d\du{\db}{j}}\hat
C\ud{ai\db}{j}\cor{d_{ai},G}
\end{gather}
Here $\hat C\ud{ai\db}{j}$ is the matrix $\hat
C\ud{ai\db}{j}=\dfrac{1}{2ip^2}p_\mu\bar \sigma^{\mu\,a
\db}\delta\ud{i}{j}$ and verifies
\begin{gather}
{C_{ai\db}}^j \hat C\ud{bk\db}{j}=\delta\du{i}{i}\delta\du{a}{b}\
\ .
\end{gather}
Calculating the Dirac brackets for all coordinates and momenta the
result is \cite{Cas2},
\begin{gather}
\dir{\theta^{ai},\theta^{bj}}=\dir{\pi_{ai},\pi_{bj}}=\dir{p_\mu,p_\nu}=0\label{con1}\\
\dir{\theta^{ai},\bar\theta\ud{\da}{j}}=\frac{-1}{2ip^2}p_\mu\bar \sigma^{\mu\,\da a}\delta^i_j\\
\dir{x^\mu,\theta^{ai}}=\frac{1}{2p^2}p_\nu \bar\sigma^{\nu\,\da a}\theta^{bi} \sigma\ud{\mu}{b\da}\\
\dir{x^\mu,\bar\theta\ud{\da}{i}}=\frac{1}{2p^2}p_\nu \bar \sigma^{\nu\,\da a}
\bar\theta\ud{\db}{i}\sigma\ud{\mu}{a\db}\label{con3}\\
\dir{x^\mu,x^\nu}=\frac{-S^\mn}{p^2}\label{con2}\ \ ,
\end{gather}
Using (\ref{d1}),(\ref{d2}) and (\ref{Sigmamna}) the equation for
the internal angular momentum $S^\mn$ (\ref{ClassicalInternal})
may be  written also in the form,
\begin{gather}
S^\mn=\ep^{\mn\rho\lambda}p_\rho\theta^{ai}\sigma_{\lambda\,
a\db}\bar \theta\ud{\db}{i}\ \ .
\end{gather}

Before applying the quantization rules to the algebra above to
construct the quantum theory of the superparticle  one can first
use a different approach which in particular identifies the
Hilbert space associate to the action (\ref{SuperAction}) . The
second class constraints are separated in two subsets, the subset
$d_a$ and its complex conjugate $\bar d_\da$. It has been proposed
that in such a situation,  if it is too complicated to impose all
the constraints it may be sufficient to impose only one of the two
sets because then the matrix elements are zero. That is in
general, for constraints $(\phi_\alpha,\bar\phi_\alpha)$ imposing
 $\hat\phi_\alpha\bra{V}=0$ then
$\ket{V}\hat{\bar\phi}_\alpha=0$.  For the superparticle, this
idea goes back to Casalbuoni\cite{Cas2} and has been applied
extensively by Lusana\cite{Lus}, Frydryszak \cite{Fry} and
collaborators. Although the application of this approach in the
cases  considered by these authors provides the correct answer,
the approach fails to work in general. When the method works
correctly it can be understood as follows. For
$\cor{\phi_\alpha,\bar\phi_\beta}$ a set of second class
constraints the measure in the corresponding functional integral
is $\left[\det\cor{\phi,\bar\phi}\right]^{1/2}$. In the particular
case in which $\det\cor{\phi_\alpha,\phi_\beta}=0$ then the above
measure reduces to $\det\cor{\phi_\alpha,\bar\phi_\beta}$ which is
exactly the functional measure for a set of first class
constraints $\phi_\alpha$ with a gauge fixing condition
\cite{GRS,RS}. In the case under study it is easily seen that
$\bar d_\da$ alone are a set of first class constraints and thus
the model is equivalent to a gauge system in which only this set
of first class constraints exists. We are then free to choose a
different gauge fixing condition. This method was applied in
reference \cite{HatResSte} to show that the wave functions of the
$D=9$, $N=2$, massive  superparticle with central charges expand a
KKB ultrashort multiplet \cite{Nicolai}.

The canonical coordinates necessary to develop the quantum theory
 are suggested by the discussion above. In the initial Lagrangean
 we change coordinates
\begin{gather}
x^\mu_L=x^\mu+i\theta\ud{a}{i}\sigma\ud{\mu}{a\db}\bar\theta^{\db
i}
\end{gather}
We can then write
\begin{gather}
\omega^\mu=\dot
x^\mu_L-2i\dot\theta\ud{a}{i}\sigma\ud{\mu}{a\db}\bar\theta^{\db
i}\ \ .
\end{gather}
The new momenta associated to the coordinates
$(x^\mu_I,\theta^a,\bar\theta^\da)$ are
\begin{gather}
p_\mu^L=e^{-1}\omega_\mu\\
\pi^L_{ai}=-2ip^L_\mu\sigma\ud{\mu}{a\da}\bar\theta\ud{\da}{i}\label{second}\\
{\bar\pi}\udu{L}{\da}{i}=0\ \ .\label{first}
\end{gather}
We may also define
\begin{gather}
d^L_{ai}=\pi^L_{ai}+2ip^L_\mu\sigma\ud{\mu}{a\da}\bar\theta\ud{\da}{i}\\
\bar d\udu{L}{\da}{i}=\bar \pi\udu{L}{\da}{i}
\end{gather}
and we have,
\begin{gather}
\cor{d^L_{ai},\bar
d\udu{L}{\da}{j}}=\delta\du{i}{j}2i\sigma\ud{\mu}{a\da}p^L_\mu={C_{ai\da}}^j\
\ .
\end{gather}
The Dirac brackets are then given by
\begin{gather}
\dir{x^\mu_L,x^\nu_L}=\dir{x^\mu_L,\theta^{ai}}=\dir{\theta^{ai},\theta^{bj}}=0\label{coorLeft1}\\
\dir{x^\mu_L,p^L_\nu}=\delta^\mu_\nu\\
\dir{\pi^L_{ai},\theta^{bj}}=-\delta\du{i}{j}\delta\du{a}{b}\label{coorLeft2}\ \ .
\end{gather}
The resulting Hilbert space is the set of chiral superfields.
The anti-chiral sector may be obtained  along the same lines.\\

\section{The quantum  algebra}

The chiral variables $X_\mu^L$ parameterize the physical degrees
of freedom of the superparticle and give a solution for the
classical dynamic problem, but to formulate the quantum theory of
this system one has still to represent the Dirac algebra
(\ref{con1})-(\ref{con2}) in terms of a set of operators in terms
of Heisenberg commutators following the rule
$\dir{\cdot,\cdot}\rightarrow-i\con{\cdot,\cdot}$ or
$-i\cor{\cdot,\cdot}$. (We use $\con{\cdot,\cdot}$ for commutators
and  $\cor{\cdot,\cdot}$ for anticommutators since no confusion
can arise with the classical brackets.). The set of chiral
superfields has been already identified as one Hilbert space where
the quantization program may be pursued. As we discuss below this
choice is not unique and when due care are given to some details
in the algebra the space of anti-chiral and the space of
superfields associated to the irreducible tensor multiplet may
also be considered.

The other problem that has to be faced, as was first noted by
Casalbuoni, is to deal with the fact that the coordinates $x^\mu$
cannot represented multiplicatively because they do not commute.
Taking this into account, our task is to find a set of operators
$\hat X^\mu$, $\hat \Theta^a$ and $\bar \Theta^\da$ that satisfy
the following algebra to which we will refer as the quantum
algebra of the superparticle,

\begin{gather}
\con{\hat P_\mu,\hat\Theta^a}=\con{\hat P_\mu,\hat
{\bar\Theta}^\da}=\cor{\hat\Theta^a,\hat\Theta^b}
=\cor{\hat{\bar\Theta}^\da,\hat{\bar\Theta}^\db}=0\label{conPrime}\\
\con{\hat X^\mu,\hat P^\nu}=i\eta^\mn\\
\cor{\hat\Theta^{a},\hat{\bar\Theta}^{\da}}=\frac{-1}{2\hat P^2}
\hat P_\mu\bar \sigma^{\mu\,\da a}\label{con1Prime}\\
\con{ \hat X^\mu, \hat\Theta^{a}}=\frac{i}{2\hat P^2}
\hat P_\nu \bar\sigma^{\nu\,\da a} \hat \Theta^b \sigma\ud{\mu}{b\da}\\
\con{ \hat X^\mu,\hat{\bar\Theta}^{\da}}=\frac{i}{2\hat P^2}\hat
P_\nu \bar \sigma^{\nu\,\da a}
\hat{\bar\Theta}^{\db}\sigma\ud{\mu}{a\db}\\
\con{ \hat X^\mu, \hat X^\nu}=\frac{-i\hat S^\mn}{\hat P^2}\ \ .
\label{ConCoord}
\end{gather}
At the quantum level we may also define
\begin{gather}
\hat\Pi_a=-i\hat
P_\mu\sigma\ud{\mu}{a\da}\hat{\bar\Theta}^\da\qquad
\hat{\bar\Pi}_\da=-i\hat P_\mu\sigma\ud{\mu}{a\da}\hat\Theta^a
\end{gather}
and consequently all commutators involving this quantities could
be obtained from this relations. The operator $\hat S^\mn$ that
appears on the right hand side of equation (\ref{ConCoord}) is the
internal angular momentum corresponding to the representation
defined by our operators. That is we should have that
\begin{gather}
\hat J^\mn=\hat X^\mu\hat P^\nu-\hat X^\mu \hat P^\nu+\hat S^\mn
\end{gather}
together with $\hat Q_a$, $\hat {\bar Q}_\da$ and $\hat P_\mu$
satisfy the super Poincar{\`e} algebra.

In the functional space of all the wave functions defined on the
Salam and Strathdee superspace  the super Poincar{\`e} algebra is
represented by the operators,
\begin{gather}
P_\mu=-i\p_\mu\\
\Pi_a=-i\p_a\qquad \bar\Pi_\da=-i\bar\p_\da\\
J_\mn=L_\mn+S_\mn=X_\mu P_\nu-X_\mu P_\nu-\frac{1}{4}
(\Theta\sigma_\mn\Pi+\bar\Pi\bar\sigma_\mn\bar\Theta)\\
Q_a=\Pi_a-iP_\mu\sigma\ud{\mu}{a\da}\bar\Theta^\da\\
\bar Q_\da=-\bar\Pi_\da+iP_\mu\sigma\ud{\mu}{a\da}\Theta^a\ \ .
\end{gather}
The operators $X_\mu$, $\Theta^a$, and $\bar\Theta^\da$  act
multiplicatively (see (\ref{Multi})).  This representation of the
super Poincar{\`e} algebra is not irreducible. Even after imposing the
mass shell condition $(P^2+m^2)\Psi=0$ the resulting
representation is reducible. Indeed it holds the following
theorem.(\cite{Oku}, \cite{RitSok},
\cite{GatGriRoeSie})\\
\textbf{Theorem: }\\
\textit{Let $\Psi(x^\mu,\theta^a,\bar\theta^\da)$ be a superfield
that satisfies
\begin{gather}
(P^2+m^2)\Psi=0\ \ ,
\end{gather}
then it can be written in a unique way ($m\neq 0$) as
\begin{gather}
\Psi=\Psi_\CC+\Psi_A+\Psi_\TT\\
\Psi_\CC=\Pcm\Psi\qquad \Psi_A=\Pam \Psi\qquad \Psi_\TT=\Ptm\Psi
\end{gather}
where the projector operators are
\begin{gather}
\Pam =\frac{1}{16m^2}D^2\bar D^2\qquad \Pcm=\frac{1}{16 m^2} \bar
D^2 D^2\qquad \Ptm=\frac{-1}{8m^2}\bar D_\da D^2\bar D^\da\ \ .
\end{gather}}
\noindent The operators representing the super Poincar{\`e} algebra
commute with any of this projectors, meaning each of the subspaces
bear a representation of the group which correspond respectively
to the chiral, anti-chiral and irreducible tensor multiplets. In
these subspaces the corresponding generators are obtained in the
form $\Proy_G J_\mn\Proy_G$ and $\Proy_G P_\mu\Proy_G$ with
$\Proy_G$ the corresponding operator.

 The observation above led us to construct a representation of
the quantum algebra of the superparticle by considering operators
acting on the subspaces of the superfields space defined by  the
chiral (C), anti-chiral (A) or tensorial (T) projectors. The
restricted operators are defined by defined by
\begin{gather}
X^\mu_G\equiv \Proy_G X^\mu \Proy_G\\
\Theta^a_G\equiv \Proy_G\Theta^a \Proy_G\\
{\bar\Theta}_G^\da\equiv \Proy_G\bar\Theta^\da \Proy_G
\end{gather}
with $\Proy_G$  any one of the projectors,
\begin{gather}
\Pa =\frac{-1}{16P^2}D^2\bar D^2, \qquad\  \Pc=\frac{-1}{16
P^2}\bar D^2 D^2,\qquad\  \Pt=\frac{1}{8P^2}\bar D_\da D^2\bar
D^\da\ \ .
\end{gather}
Note that the projectors  have been taken out of the mass shell.

Below we show explicitly that each set of operators satisfy the
quantum algebra (\ref{conPrime}-\ref{ConCoord}). In each of the
subspaces the operator which we denote in generic form as $\hat
S^\mn$ at the right hand side of (\ref{ConCoord}) results to be
the projected internal angular momentum in the corresponding
subspace. We introduce the notation, $S^\mn_G\equiv \Proy_G
S^\mn\Proy_G$ for these operators.

We begin discussing the representation acting on the chiral
superfields which after the discussion at the final of the
previous section should correspond to the quantum theory of the
superparticle defined by the classical action (\ref{SuperAction}).
We present the result in the form of the following theorem.\\
\textbf{Theorem}\\
\textit{The set of operators defined by $X^\mu_\CC=\Pc X^\mu\Pc$,
$\Theta^a_\CC=\Pc\Theta^a\Pc$ and
$\bar\Theta^\da_\CC=\Pc\bar\Theta^\da\Pc$ satisfy the algebra
(\ref{conPrime}-\ref{ConCoord}). The operator $S^\mn_\CC\equiv \Pc
S^\mn\Pc$ can be written as
\begin{gather}
S^\mn_\CC=-\frac{1}{4}\left(\Theta_\CC\sigma^\mn\Pi_\CC+
\bar\Pi_\CC\bar\sigma^\mn\bar \Theta_\CC\right)=
\ep^{\mn\rho\lambda}P_\rho\Theta^{a}_\CC\sigma_{\lambda\,a\db}
\bar\Theta^{\db}_\CC
\end{gather}}.\\
\textbf{Proof}\\
To compute the anticommutator $\cor{\Theta_\CC^a,\Theta_\CC^b}$,
substitute $\Theta^a_\CC$ by $\Pc\Theta^a\Pc$,
\begin{gather}
\cor{\Theta_\CC^a,\Theta_\CC^b}=(\Pc\Theta^a\Pc)(\Pc\Theta^b\Pc)+(\Pc\Theta^b\Pc)(\Pc\Theta^a\Pc)
\end{gather}
and use the fact that $\Pc^2=\Pc$. Then note that
\begin{gather}
\Pc\Theta^a\Pc\Theta^b\Pc=\Pc\Theta^a\con{\Pc,\Theta^b}\Pc+\Pc\Theta^a\Theta^b\Pc=\Pc\Theta^a\Theta^b\Pc
\end{gather}
With the help of (\ref{Pctheta}) observe that
$\con{\Pc,\Theta^b}\Pc=0$. Then,
\begin{gather}
\cor{\Theta^a_\CC,\Theta^b_\CC}=\Pc(\Theta^a\Theta^b+\Theta^b\Theta^a)\Pc=0
\end{gather}
The anticommutator $\cor{\bar\Theta_\CC^\da,\bar\Theta_\CC^\db}=0$
is also straightforward. To address the first non zero commutator:
\begin{gather}
\cor{\Theta_\CC^a,{\bar\Theta_\CC}^\da}=\Pc\Theta^a\Pc\bar\Theta^\da
\Pc+\Pc\bar\Theta^\da \Pc\Theta^a\Pc\ \ .
\end{gather}
use again the formula (\ref{Pctheta}) to prove that
\begin{gather}
\cor{\Theta_\CC^a,{\bar\Theta_\CC}^\da}=\Pc\con{\Theta^a,\Pc}\bar\Theta^\da
\Pc+\Pc\bar\Theta^\da\con{\Pc,\Theta^a}\Pc=\nonumber\\
\frac{-1}{8P^2}\Pc\left(-\bar
D^2D^a\bar\Theta^\da+\bar\Theta^\da\bar D^2D^a\right)
\Pc=\nonumber\\
\frac{-1}{8P^2}\Pc\left(-\con{\bar
D^2,D^a}\bar\Theta^\da+\bar\Theta^\da
\con{\bar D^2,D^a}\right)\Pc=\nonumber\\
\frac{-1}{8P^2}\Pc\left(4i\bar\sigma^{\mu\,\db a}\bar
D_\db\p_\mu\bar\Theta^\da+\bar\Theta^\da(-4i\bar\sigma^{\mu\,\db
a}\bar D_\db\p_\mu)\right)\Pc=\nonumber\\
\Pc\left(\frac{4i}{8P^2}\bar\sigma^{\mu\,\da a}\p_\mu\right)
\Pc=-\frac{1}{2P^2}\bar\sigma^{\mu\,\da a}P_\mu \ . \label{chi1}
\end{gather}
The next task is to compute $\con{X_\CC^\mu,\Theta_\CC^a}$. The
tricks to be used are similar:
\begin{gather}
\con{ X_\CC^\mu, \Theta_\CC^a}=\Pc X^\mu \Pc \Theta^a \Pc
-\Pc\Theta^a\Pc X^\mu \Pc=\Pc X^\mu \con{\Pc, \Theta^a} \Pc
-\Pc\con{\Theta^a,\Pc} X^\mu \Pc\ \ .
\end{gather}
Recalling once again that $\con{\Pc,\Theta^a}\Pc=0$,
\begin{gather}
\con{ X_\CC^\mu, \Theta_\CC^a}=\frac{-1}{8P^2}\Pc \bar D^2 D^a
X^\mu\Pc=
\frac{-1}{8P^2}\Pc \con{\bar D^2,D^a} X^\mu\Pc=\nonumber\\
\frac{4i}{8P^2}\Pc\bar\sigma^{\nu\,\db a}\p_\nu \bar D_\db
X^\mu\Pc=\Pc\left(\frac{iP_\nu}{P^2}\Theta^b\sigma\ud{\mu}{b\db}\bar\sigma^{\nu\,\db
a}\right)\Pc\ \ . \label{chi2}
\end{gather}
The next commutator is $ \con{ X_\CC^\mu, X_\CC^\nu}$ and is the
most involved
\begin{gather}
\con{ X_\CC^\mu, X_\CC^\nu}=\Pc X^\mu \Pc X^\nu \Pc-\Pc X^\nu \Pc
X^\mu \Pc=\nonumber\\
=\Pc \con{X^\mu, \Pc} X^\nu \Pc-\Pc \con{X^\nu, \Pc} X^\mu \Pc \ .
\end{gather}
Using formulas (\ref{xOverP}, \ref{DsqareX}) and
(\ref{BarDsqareX}),  we can write
\begin{gather}
\con{X^\mu,\Pc}=-\frac{1}{P^2}\con{X^\mu,\bar D^2D^2}-\frac{1}{16}\con{X^\mu,\frac{1}{P^2}}\bar D^2 D^2=\nonumber\\
=\frac{-1}{16P^2}\left(\bar D^2\con{X^\mu,D^2}+\con{X^\mu,\bar D^2}D^2\right)-\frac{i}{8}\frac{P^\mu}{P^2}\frac{1}{P^2}\bar D^2 D^2=\nonumber\\
=\frac{i}{8}\left(\bar
D^2(D\sigma^\mu\bar\Theta)-(\Theta\sigma^\mu\bar
D)D^2\right)+\frac{2iP^\mu}{P^2}\Pc \ .
\end{gather}
With this computation the commutator is
\begin{gather}
\con{ X_\CC^\mu, X_\CC^\nu}=\frac{i}{8P^2}\Pc\left(\bar D^2(D\sigma^\mu\bar\Theta)X^\nu- (\Theta\sigma^\mu\bar D)D^2X^\nu+(\mu\leftrightarrow\nu)\right)\Pc-\nonumber\\
-\frac{2i}{P^2}\Pc\left(X^\mu P^\nu-X^\nu P^\mu\right)\Pc
\end{gather}
We recognize a term proportional to $L^\mn_\CC$. To further
simplify the other term note that $\Pc\bar D^2
D^a=\Pc\con{D^2,D^a}$ and $\Pc\Theta^a\bar D^\da
D^2=\Pc\Theta^a\con{\bar D^\da, D^2}$. Using formulas
(\ref{DsquareDa}) and (\ref{DaDsquare}) the commutator reads
\begin{gather}
\con{X^\mu_\CC,X^\nu_\CC}=\frac{2i}{P^2}L^\mn_\CC+\nonumber\\
+\frac{i(-4i)}{8P^2}\Pc\left(\bar D\bar\sigma^\lambda\sigma^\mu\bar \Theta\p_\lambda X^\nu-\Theta\sigma^\mu\bar\sigma^\lambda D\p_\lambda X^\nu-(\mu\leftrightarrow\nu)\right)\Pc\nonumber\\
\con{X^\mu_\CC,X^\nu_\CC}=\frac{2i}{P^2}L^\mn_\CC+\frac{1}{2P^2}(I_1^\mn+I_2^\mn)
\end{gather}
To disentangle this equation first write $\p_\lambda
X^\mu=\delta\du{\lambda}{\mu}+X^\mu\p_\lambda$ so that
\begin{gather}
I_1^\mn=\Pc\left(\bar D\bar\sigma^\lambda\sigma^\mu\bar
\Theta\p_\lambda X^\nu-
(\mu\leftrightarrow\nu)\right)\Pc=\nonumber\\
\Pc\left(\bar
D\bar\sigma^\nu\sigma^\mu\bar\Theta-\Tr(\bar\sigma^\lambda\sigma^\mu)X^\nu\p_\lambda-
(\bar\sigma^\lambda\sigma^\mu)\ud{\da}{\db}\bar\Theta^\db(\bar
D_\da X^\nu)\p_\lambda- (\mu\leftrightarrow\nu)\right)\Pc
\end{gather}
The first of this terms is 0, $\Pc \bar
D\bar\sigma^\mn\bar\Theta\Pc=0$ because $\Tr(\sigma^\mn)=0$. The
second is $2\Pc(X^\nu \p^\mu-X^\mu\p^\nu)\Pc=-2i L_\CC^{\mn}$. The
last piece is
\begin{gather}
\Pc\left(i(\bar\sigma^\lambda\sigma^\mu)\ud{\da}{\db}\bar\Theta^\db\sigma\ud{\nu}{b\da}\Theta^a\p_\lambda-
(\mu\leftrightarrow\nu)\right)\Pc=\Pc(\Theta(\sigma^\nu\bar\sigma^\lambda\sigma^\mu-
\sigma^\mu\bar\sigma^\lambda\sigma^\nu)\bar\Theta P_\lambda)\Pc
\end{gather}
Using the  identity (\ref{AntiBasicSigma})  $I_1^\mn$ is given by
\begin{gather}
I_1^\mn=-2iL_\CC^\mn-2i\ep^{\mn\alpha\lambda}\Pc\Theta\sigma_\lambda\bar\Theta\Pc
\end{gather}
The other term, $I_2^\mn$ is simpler
\begin{gather}
I_2^\mn=-\Pc\left(\Theta\sigma^\mu\bar\sigma^\lambda D\p_\lambda
X^\nu-
(\mu\leftrightarrow\nu)\right)\Pc=\nonumber\\
=-\Pc\left((\sigma^\mu\bar\sigma^\lambda)\du{a}{b}\cor{\Theta^a,D_b}\p_\lambda
X^\nu-
(\mu\leftrightarrow\nu)\right)\Pc=\nonumber\\
=2\Pc(\p^\mu X^\nu-\p^\nu X^\mu)\Pc=-2iL^\mn_\CC
\end{gather}
Taking all pieces together
\begin{gather}
\con{X^\mu_\CC, X^\nu_\CC}= \frac{2i}{P^2}L_\CC^\mn+
\frac{1}{2P^2} \left(-2iL_\CC^\mn-
2i\ep^{\mn\alpha\lambda}\Pc\Theta\sigma_\lambda\bar\Theta\Pc-2iL_\CC^\mn\right)
\end{gather}
Summing up we get,
\begin{gather}
\label{chi3} \con{ X_\CC^\mu,
X_\CC^{\nu}}=\frac{-i}{P^2}S_\CC^\mn\ \ .
\end{gather}
In this case $S^\mn_\CC$ can be written in terms of the
corresponding projected operators. To this end note that
$\Pc\bar\Theta^\da\Theta^a\Pc=\Pc\bar\Theta^\da\Pc\Theta^a\Pc$.
since $\con{\Theta^a,\Pc}\Pc=0$. We have then,
\begin{gather}
 S^\mn_\CC=\Pc S^\mn
\Pc=\Pc\ep^{\mn\rho\lambda}P_\rho\Theta^{a}\sigma_{\lambda\,a\db}
\bar\Theta^{\db}\Pc=
\ep^{\mn\rho\lambda}P_\rho\Theta_\CC^{a}\sigma_{\lambda\,a\db}
\bar\Theta_\CC^{\db}\label{Internal}
\end{gather}
which satisfies  Pryce's constraint $P_\mu S^{\mn}_\CC=0$
\\
This end the proof of the theorem.

The commutators (\ref{chi1}-\ref{chi2},\ref{chi3}) realize the
quantum algebra of the standard massive superparticle and hence we
have a complete solution for the quantization of this system. The
computations with the anti-chiral projector are completely
analogous and led to exactly the same algebra. This give an
equivalent but different covariant solution to the quantization of
the superparticle.

We consider now the representation of the algebra using the
tensorial projector.\\
\textbf{Theorem}\\
\textit{The set of operators defined by $X^\mu_\TT=\Pt X^\mu\Pt$,
$\Theta^a_\TT=\Pt\Theta^a\Pt$ and
$\bar\Theta^\da_\TT=\Pt\bar\Theta^\da\Pt$ satisfy the algebra
(\ref{conPrime}-\ref{ConCoord}). The operator $S^\mn_\TT\equiv \Pt
S^\mn\Pt$ can be written as
\begin{gather}
S^\mn_\TT=\tilde
S^\mn_\TT+\frac{P_\alpha}{4P^2}\ep^{\mn\alpha\lambda}\Pt\bar
D\bar\sigma_\lambda D\Pt
\end{gather}}
with
\begin{gather}
\label{tilde} \tilde
S^\mn_\TT=-\frac{1}{4}\left(\Theta_\TT\sigma^\mn\Pi_\TT+\bar\Pi_\TT\bar\sigma^\mn\bar\Theta_\TT\right)
\end{gather}
\textbf{Proof}\\
The computation of $\cor{\Theta_\TT^a,\Theta_\TT^b}=0$ and
$\cor{\bar\Theta_\TT^\da,\bar\Theta_\TT^\db}=0$ are again
straightforward. Consider,
\begin{gather}
\cor{\Theta_\TT^a,{\bar\Theta_\TT}^\da}=\Pt\Theta^a\Pt\bar\Theta^\da
\Pt+\Pt\bar\Theta^\da \Pt\Theta^a\Pt\ .
\end{gather}
Taking advantage of (\ref{Pttheta}) and $\bar D^2 \Pt=\Pt \bar
D^2=0$,
\begin{gather}
\cor{\Theta_\TT^a,{\bar\Theta_\TT}^\da}=\Pt\con{\Theta^a,\Pt}\bar\Theta^\da
\Pt+\Pt\bar\Theta^\da\con{\Pt,\Theta^a}\Pt=\nonumber\\
\frac{1}{8P^2}\Pt(-\cor{D^a,\bar D^2}\bar \Theta^\da
+\bar\Theta^\da\cor{\bar D^2, D^a})\Pt=\nonumber\\
\frac{1}{8P^2}\Pt(-\con{D^a,\bar
D^2}\bar\Theta^\da+\bar\Theta^\da\con{\bar D^2, D^a})\Pt\ \ .
\end{gather}
Recalling  again  (\ref{DsquareDa}),
\begin{gather}
\label{tens1}
\cor{\Theta_\TT^a,{\bar\Theta_\TT}^\da}=\frac{-4i\bar\sigma^{\mu\,\db
a}}{8P^2}\Pt\left(\bar D_\db\bar\Theta^\da+\bar\Theta^\da\bar
D_\db\right)\Pt\p_\mu=
\Pt\left(\frac{4i}{8P^2}\bar\sigma^{\mu\,\da a}\p_\mu\right) \Pt
\end{gather}
Next  compute,
\begin{gather}
\con{ X_\TT^\mu, \Theta_\TT^a}= \Pt X^\mu \Pt \Theta^a
\Pt-\Pt\Theta^a \Pt X^\mu \Pt=
\Pt X^\mu\con{\Pt,\Theta^a}\Pt-\Pt\con{\Theta^a,\Pt}X^\mu \Pt=\nonumber\\
\frac{1}{8P^2}\Pt\left(X^\mu\con{\bar D^2,D^a}+\con{D^a,\bar D^2}X^\mu\right)\Pt=\nonumber\\
\frac{1}{8P^2}\Pt\left(-X^\mu 4i\bar\sigma^{\nu\,\db a}\bar
D_\db\p_\nu+4i\bar\sigma^{\nu\,
\db a}\bar D_\db\p_\nu X^\mu\right)\Pt=\nonumber\\
\frac{\bar\sigma^{\nu\,\db a}}{2iP^2}\Pt(X^\mu\bar D_\db-\bar
D_\db
X^\mu)\Pt\p_\nu=\Pt\left(\frac{i}{2P^2}\Theta^b\sigma\ud{\mu}{b\db}\bar\sigma^{\nu\,\db
a}\right)\Pt P_\nu\ \ . \label{tens2}
\end{gather}
Finally consider,
\begin{gather}
\con{ X_\TT^\mu, X_\TT^\nu}=\Pt X^\mu \Pt X^\nu \Pt-\Pt X^\nu\Pt
X^\mu\Pt=\Pt\con{X^\mu,\Pt}X^\nu\Pt -\Pt\con{X^\nu,\Pt}X^\mu\Pt
\end{gather}
Taking into account the  previous computation
\begin{gather}
\Pt\con{X^\mu,\Pt}=-\Pt\con{X^\mu,\Pc}-\Pt\con{X^\mu,\Pa}=\nonumber\\
\frac{-1}{16P^2}\Pt(\bar D^2\con{D^2,X^\mu}+\con{\bar D^2,X^\mu}
D^2+D^2\con{\bar D^2, X^\mu}+
\con{D^2,X^\mu}\bar D^2)=\nonumber\\
\frac{-1}{16P^2}\Pt(\con{\bar D^2,X^\mu}D^2+\con{D^2,X^\mu}\bar
D^2)
\end{gather}
Introducing this value in the commutator and using (\ref{DsqareX})
and (\ref{BarDsqareX})
\begin{gather}
\label{dline} \con{ X_\TT^\mu, X_\TT^\nu}=\frac{2i}{16
P^2}\Pt(D\sigma^\mu\bar\Theta\bar D^2-
\Theta\sigma^\mu\bar D D^2)X^\nu\Pt-(\mu\leftrightarrow\nu)=\nonumber\\
\frac{i}{8P^2}\Pt(D\sigma^\mu\bar \Theta\con{\bar D^2,X^\nu}-
\Theta\sigma^\mu\bar D\con{D^2,X^\nu})\Pt-(\mu\leftrightarrow\nu)=\nonumber\\
\frac{1}{4P^2}\Pt(D\sigma^\mu\bar\Theta\Theta\sigma^\nu\bar
D+\Theta\sigma^\mu\bar
DD\sigma^\nu\bar\Theta)\Pt-(\mu\leftrightarrow\nu)
\end{gather}
The  last two terms in (\ref{dline}) can be written as
\begin{gather}
\frac{1}{4P^2}\Pt \left(\con{D\sigma^\nu\bar\Theta,
\Theta\sigma^\nu \bar D}-(\mu\leftrightarrow\nu)\right)\Pt
\end{gather}
The commutator here is given by
\begin{gather}
\con{D^a\sigma\ud{\nu}{a\da}\bar\Theta^\da,\Theta^b\sigma\ud{\mu}{b\db}\bar
D^\db}- (\mu\leftrightarrow\nu)=
\sigma\ud{\nu}{a\da}\sigma\ud{\mu}{b\db}\con{D^a\bar\Theta^\da,
\Theta^b\bar D^\db}-(\mu\leftrightarrow\nu)=\nonumber\\
\sigma\ud{\nu}{a\da}\sigma\ud{\mu}{b\db}\left(D^a\con{\bar\Theta^\da,\Theta^b\bar
D^\db}+
\con{D^a,\Theta^b\bar D^\db}\bar\Theta^\da\right)-(\mu\leftrightarrow\nu)=\nonumber\\
\sigma\ud{\nu}{a\da}\sigma\ud{\mu}{b\db}\left(-D^a\Theta^b\cor{\bar\Theta^\da,\bar
D^\db}+ \cor{D^a,\Theta^b}\bar
D^\db\bar\Theta^\da-\Theta^b\cor{D^a,\bar
D^\db}\bar\Theta^\da\right)-
(\mu\leftrightarrow\nu)=\nonumber\\
\sigma\ud{\nu}{a\da}\sigma\ud{\mu}{b\db}(-D^a\Theta^b\ep^{\da\db}+\ep^{ab}\bar
D^\db\bar\Theta^\da-2P_\alpha\bar\sigma^{\alpha\,\db
a}\Theta^b\bar\Theta^\da)-(\mu\leftrightarrow\nu)\label{Long}
\end{gather}
The first term is
\begin{gather}
\sigma\ud{\nu}{a\da}\sigma\ud{\mu}{b\db}(-D^a\Theta^b\ep^{\da\db})-(\mu\leftrightarrow\nu)=
-\ep^{\da\db}\sigma\ud{\nu}{a\da}\sigma\ud{\mu}{b\db}\ep^{ac}D_c\Theta^b-
(\mu\leftrightarrow\nu)=\nonumber\\
-{\sigma^\mn}\du{b}{c}D_c\Theta^b=\Theta\sigma^\mn D
\end{gather}
using that the matrix $\sigma^\mn$ is traceless. A similar
computation for the second term in equation (\ref{Long}) yields
the result $ -\bar D\bar\sigma^\mn\bar\Theta $. With the aid of
(\ref{AntiBasicSigma}) the third and last term sum up to,
\begin{gather}
-2P_\alpha\Theta\left(\sigma^\nu\bar\sigma^\alpha\sigma^\mu-
\sigma^\mu\bar\sigma^\alpha\sigma^\nu\right)\bar\Theta=
4iP_\alpha\ep^{\nu\alpha\mu\lambda}\Theta\sigma_\lambda\bar\Theta
\end{gather}
and taking all the terms together ,
\begin{gather}
\con{ X_\TT^\mu, X_\TT^\nu}=\frac{1}{4P^2}\Pt
\left(\Theta\sigma^\mn D-\bar
D\bar\sigma^\mn\bar\Theta+4iP_\alpha\ep^{\nu\alpha\mu\lambda}\Theta\sigma_
\lambda\bar\Theta\right)\Pt
\end{gather}
Using the explicit form of $D_a$ and $\bar D_\da$ shown  in the
appendix and formula (\ref{Sigmamna}),
\begin{gather}
\frac{1}{4P^2}\left(\Theta\sigma^\mn D-\bar
D\bar\sigma^\mn\bar\Theta+
4iP_\alpha\ep^{\nu\alpha\mu\lambda}\Theta\sigma_\lambda\bar\Theta\right)=\nonumber\\
\frac{i}{4P^2}\left(\Theta\sigma^\mn\Pi+\bar\Pi\sigma^\mn\bar\Theta\right)+
\frac{1}{4P^2}\left(-P_\alpha\Theta\left(\sigma^\mn\sigma^\alpha+
\sigma^\alpha\sigma^\mn\right)\bar\Theta+
4iP_\alpha\ep^{\nu\alpha\mu\lambda}\Theta\sigma_\lambda\bar\Theta\right)= \nonumber\\
\frac{-iS^\mn}{P^2}+\frac{P_\alpha}{4P^2}(-4i\ep^{\mn\alpha\lambda}+
4i\ep^{\nu\alpha\mu\lambda})
\Theta\sigma_\lambda\bar\Theta=\frac{-iS^\mn}{P^2} \ \ .
\end{gather}
Finally,
\begin{gather}
\con{ X_\TT^\mu, X_\TT^{\nu}}=\frac{-i}{P^2}
S_\TT^\mn\label{tens3}
\end{gather}
with $ S_\TT^\mn=\Pt S^\mn \Pt$. Note that this is not equal to
(\ref{tilde}) and therefore this set of operators do not
correspond to the quantization of (\ref{SuperAction}). The
operator $S^\mn_\TT$ is not written solely in terms of the $T$
operators. There is an extra term that accounts for the internal
super spin. We complete
this observation with the following lemma.\\
\textbf{Lemma}\\
\begin{gather}
S^\mn_\TT=\tilde
S^\mn_\TT+\frac{P_\alpha}{4P^2}\ep^{\mn\alpha\lambda}\Pt\bar
D\bar\sigma_\lambda D\Pt
\end{gather}
Note also that in the chiral (or anti-chiral) case,
\begin{gather}
P_\alpha\ep^{\mn\alpha\lambda}\Pc\bar D\bar\sigma_\lambda D\Pc=0
\end{gather}\\
\textbf{Proof}\\
To start insert $I=\Pt+\Pc+\Pa$ in the formula for $S^\mn_\TT$
\begin{gather}
S^\mn_\TT=
\frac{-1}{4}\Pt\left(\Theta\sigma^\mn(\Pt+\Pc+\Pa)\Pi+\bar\Pi\bar\sigma^\mn(\Pt+\Pc+
\Pa)\bar\Theta\right)\Pt= \nonumber\\
=\tilde
S^\mn_\TT+\frac{-1}{4}\Pt\left(\Theta\sigma^\mn(\Pc+\Pa)\Pi+\bar\Pi\bar\sigma^\mn(\Pc+
\Pa)\bar\Theta\right)\Pt
\end{gather}
Note that due to (\ref{Pctheta}) the terms with $\Pc$ disappear.
Then with the aid of,
\begin{gather}
\Delta S^\mn_\TT=S^\mn_\TT-\tilde
S^\mn_\TT=\frac{-1}{4}\Pt\left(\con{\Theta^a,\Pa}\Pi_b\sigma\udu{\mn}{a}{b}+
\bar\sigma\ud{\mn\,\da}{\db}\bar \Pi_\da\con{\Pa,\bar\Theta^\db}\right)\Pt=\nonumber\\
=\frac{-1}{4}\Pt\left(\frac{1}{8P^2}D^a\bar
D^2\Pi_b\sigma\udu{\mn}{a}{b}-
\frac{1}{8P^2} \bar\sigma\ud{\mn\,\da}{\db}\bar \Pi_\da D^2\bar D^\db\right)\Pt=\nonumber\\
=\frac{-iP_\alpha}{16
P^2}\Pt\left[D(\sigma^\mn\sigma^\alpha+\sigma^\alpha\bar\sigma^\mn)\bar
D\right]\Pt
=\nonumber\\
=\frac{-P_\alpha}{4P^2}\ep^{\mn\alpha\lambda}\Pt D\sigma_\lambda
\bar D\Pt
\end{gather}
This last formula completes the proof of the lemma and the theorem.\\
 The subspaces defined by the chiral and the tensor projectors bear irreducible representations of $N=1$
super Poincar{\`e} with superspin $0$ and $1/2$ respectively. As final
observation before closing this section let us compute the
relevant Casimir operator for each of the representations we have
obtained. We first define the analog of the Pauli-Lubanski four
vector
\begin{gather}
W_\mu=\frac{1}{2}\ep_{\mn\alpha\beta}\hat P^\nu\hat
J^{\alpha\beta}+ \frac{1}{8}\hat {\bar Q}\bar \sigma_\mu \hat Q
\end{gather}
where $\hat P_\mu$, $\hat J^\mn$, $\hat Q_a$ and $\hat{\bar
Q}_\da$ are the projected operators. This definition is taken from
Salam and Strathdee \cite{SalStr74} or Sokatchev \cite{Sok}. As we
shall show below it is transverse in the spaces defined by the
chiral and  the tensor projectors, although it is not transverse
in the general setting (readers may wish to compare this
definition with that of \cite{Oku} which is transverse from the
beginning). With this four vector we construct a Casimir:
\begin{gather}
W=W_\mu W^\mu
\end{gather}
In a given irreducible representation $W$ is equal to $-P^2Y(Y+1)$
with $Y$ is the super spin of the representation. For the chiral
(and the anti-chiral of course) it is not difficult to show that
\begin{gather}
W_\mu^\CC=0
\end{gather}
Which implies naturally that $W_\CC=0$ as we might expect.\\
For the quantum algebra defined by means of the tensorial
projector, the non zero term came from the new term in the
expression for the internal angular momentum
\begin{gather}
W^\TT_\mu=\frac{-1}{4}\Pt\bar D\bar\sigma_\mu D\Pt-\frac{1}{2}P_\mu\Pt\\
W_\TT=\left(\frac{-1}{4}\Pt\bar D\bar\sigma_\mu D\Pt-
\frac{1}{2}P_\mu\Pt\right) \left(\frac{-1}{4}\Pt\bar
D\bar\sigma^\mu D\Pt-\frac{1}{2}P^\mu\Pt\right)=
\nonumber\\
=\Pt\left(\frac{1}{16}(\bar D\bar\sigma_\mu D)(\bar D\bar
\sigma^\mu D)+ \frac{1}{8}P_\mu\bar D\bar \sigma^\mu D\right)\Pt
\end{gather}
Now  use (\ref{BarSBarS}) and Okubo's formula (\ref{OkuDD}) to
finally obtain,
\begin{gather}
W_\TT=-P^2\frac{1}{2}\left(\frac{1}{2}+1\right)
\end{gather}

\section{The superspin $1/2$ superparticle}
In the previous section we show that the quantum  algebra of the
superspace  may be realized by projecting the observables on the
three supersymmetric sectors of the Hilbert space of superfields.
The chiral and anti-chiral realizations correspond to the usual
$D=4$ massive superparticle as we demonstrate in section
\ref{chir} using the corresponding chiral or anti-chiral
variables. We have learned that there are three kinds of
superparticles in the Salam-Strathdee superspace. It is natural to
ask which is the classical action which upon quantization give
rise to the superspin $1/2$ superparticle. This action is given by
\begin{gather}
S=\frac{1}{2}\int\left\{e^{-1}\omega^\mu\omega_\mu-e
m^2+\ell(\dot\theta^2+ \dot{\bar\theta}^2)\right\}d\tau\label{newaction}
\end{gather}
The field $\ell(\tau)$ is a new variable that  enters in the
action besides the coordinates $(x^\mu,\theta^a,\bar\theta^{\dot
a})$. This system has no second class constraints. The condition
$\pi_\ell=0$ is a first class constraint. Preservation of this
constraint implies
\begin{gather}
\dot\theta^2+\dot{\bar\theta}^2=0\ \ .
\end{gather}
We also  have,
\begin{gather}
\pi_a+ip_\mu\sg{\mu}{a\db}\dot{\bar\theta}^\db+\ep_{ab}\ell\dot\theta^b=0\
\ ,
\end{gather}
and defining
\begin{gather}
d_a=\pi_a+ip_\mu\sg{\mu}{a\db}\dot{\bar\Theta}^\db=\ep_{ab}\ell\dot\theta^b\
\ ,
\end{gather}
we end up for $\ell\neq0$ with the following first class
constraints
\begin{gather}
\pi_\ell=0\\
p^2+m^2=0\\
d^2+\bar d^2=0\ \ .
\end{gather}
The quantum mechanics of this system is now straightforward.
Imposing the constraints on a superfield
$\Psi(x^\mu,\theta^a,\bar\theta^\da)$ we have
\begin{gather}
(P^2+m^2)\Psi(x^\mu,\theta^a,\bar\theta^\da)=0\label{p2m2}\\
(D^2+\bar D^2)\Psi(x^\mu,\theta^a,\bar\theta^\da)=0\label{d2d2}
\end{gather}
which after simple algebraic manipulations led to
\begin{gather}
(P^2+m^2)\Psi=0\\
D^2\Psi=\bar D^2\Psi=0
\end{gather}
In turn this equations are equivalent to
\begin{gather}
(P^2+m^2)\Psi=0 \qquad\Pt \Psi=\Psi\label{p2d2tensor}
\end{gather}
To prove this last statements observe that if (\ref{p2m2}) and (\ref{d2d2})
hold then $D^2\bar D^2\Psi=0$ and $\bar D^2D^2\Psi=0$ which imply that
$\cor{D^2,\bar D^2}\Psi=0$. With the identity (\ref{anticonD2barD2})
we now see that (\ref{p2d2tensor}) holds which clearly imply that $D^2\Psi=\bar D^2\Psi=0$.

This establishes that the wave functions  for this system are
superfields restricted by the tensorial projector. We note that in
our formulation the Weyl spinors satisfy a Klein Gordon field
equation. This  is equivalent to a Dirac equation for the uniquely
associated Dirac spinor.

Since this is a gauge system we need not to realize the complete
algebra of the observables to extract the physical content.
Nevertheless we observe that it should exist a covariant gauge
fixing condition for which the algebra of eqs.
(\ref{tens1}),(\ref{tens2}) and (\ref{tens3}) for the quantum
phase space variables is realized.

Some comments are in order. Action (\ref{newaction}) with $\ell$
proportional to the inverse of the einbein $e(\tau)$ was first
proposed by Volkov and Pashnev in \cite{VolPas} and was also
considered in \cite{BrinkS} . The resulting spectrum of physical
states consists of the chiral , anti-chiral and irreducible tensor
multiplets with related masses. Upon quantization the
Volkov-Pashnev system leads to unitarity problems. This can be
understood in the following way. If we keep $\ell$ fixed then
there is only one first class constraint and the space of states
is the whole space of wave functions. We can prove that in this
space there is no supersymmetric positive definite inner product.
It is necessary (in order to have a real representation of
supersymmetry) that
\begin{gather}
(Q_a)^\dag=\bar Q_\da\qquad (J_\mn)^\dag=J_\mn
\end{gather}
This implies that
\begin{gather}
(\Theta^a)^\dag=\bar\Theta^\da\qquad\qquad (\Pi_a)^\dag=\bar \Pi_\da
\end{gather}
which in turn implies
\begin{gather}
(D_a)^\dag=-\bar D_\da
\end{gather}
Now consider a chiral superfield
$\Psi(x^\mu,\theta^a,\bar\theta^\da)$ and let us assume that there
is a scalar product such that $(\Psi,\Psi)=1$. We built the wave
function $D_a\Psi$. It is easily seen with the use of
\ref{interwinding} that this wave function is in the space
restricted by the tensorial projector. It satisfies,
\begin{gather}
(D_a\Psi,D_a\Psi)=-(\Psi,\bar D_\da
D_a\Psi)=-2\sigma\ud{\mu}{a\da}(\Psi,P_\mu\Psi)<0
\end{gather}
The last inequality follows from supersymmetry
\begin{gather}
0<(Q_a\Psi,Q_a\Psi)=2\sigma\ud{\mu}{a\da}(\Psi,P_\mu\Psi)
\end{gather}
This is also the conclusion obtained by Volkov and Pashnev. The
problem will always appear if one retains the whole space of
superfields. In distinction, for our system $\ell(\tau)$ acts as
an independent lagrange multiplier and we are restricted to the
sector of the superfields space associated to the irreducible
tensor multiplet which is a Hilbert space. In this case $\Psi$ and
$Q_a\Psi$ cannot be simultaneously in the chosen Hilbert space. To
sum up, it is possible to have a Hilbert space structure on the
spaces projected by the chiral, anti-chiral or tensorial
projectors but not in the whole space of superfields. For a
related discussion see \cite{const}.

\section{Projectors and second class constraints}

In this section we put in a general setup our experience with the
quantum algebra of the superparticle. Geometrically second class
constraints restrict the symplectic manifold of the phase space to
a symplectic submanifold and first class constraints further
reduce it to a foliated Poisson manifold. When second class
constraints are present, Dirac algorithm has to be handle with
care.  In the transition to quantum theory it may appear ordering
problems and locality problems related with the inverse of the
matrix of the constraints as an operator.   At the classical level
Darboux theorem assures that there exists a set of canonical
coordinates but even if this set of coordinates is obtained  it
should be considered that the  quantization may be not
straightforward. It is useful and interesting to represent
directly the Dirac algebra without dwelling in a specific choice
of independent coordinates.

Let us consider the general case of a classical system with
coordinates and momenta $(q^i,p_j)$ subject to second class
constraints $\phi_\alpha(q^i,p_j)$ satisfying the algebra
\begin{gather}
\cor{\phi_\alpha,\phi_\beta}=C_{\alpha\beta}\ \ ,
\end{gather}
with the Dirac brackets given by
\begin{gather}
\dir{F,G}=\cor{F,G}-\cor{F,\phi_\alpha}C^{\alpha\beta}\cor{\phi_\beta,G}\
\ .
\end{gather}
First we note that since second class constraints reduce the phase
space in quantum mechanics this should mean that the physical
Hilbert space is a subspace of the whole Hilbert space
$\mathcal{H}$. Any such subspace is characterized  by a projection
operator $\Proy$ with the physical Hilbert subspace given by
$\Proy \mathcal{H}$. The operators acting on the physical space
are those projected from the whole Hilbert space $\mathcal{H}$.

What we observe is that there should exist a projection operator
$\Proy_G$ such that the quantum algebra is represented on the
Hilbert subspace $G$ in a generalized Schr{\"o}dinger picture by
\begin{gather}
Q_G^i\rightarrow \Proy_GQ^i\Proy_G\\
P_{Gi}\rightarrow \Proy_G(-i\hbar\p_i)\Proy_G
\end{gather}
As we saw in the case of the superparticle this proposal provides
us with an elegant solution to the representation of  the
superparticle algebra.

\section{Conclusions and outlook}

In this paper we have presented the quantization of the massive
superparticle from new points of view which reveal some aspects of
the structure of the superspace previously unnoted. We  have
presented the canonical reduction of the classical action
(\ref{SuperAction}) to its physical degrees of freedom. This
procedure identifies in the most direct way the wave functions of
this system as the chiral superfields. We then have constructed an
explicit representation of the quantum observables of the system
acting on the space of chiral superfields. This have been achieved
by considering the projection of the  operators acting on
arbitrary superfields on the set of chiral superfields. This
procedure unravel also the possibility of having  a different but
related quantum algebra realized on the space of superfields
associated to the irreducible tensor multiplet. To complete the
physical picture we display a new classical action whose wave
functions are shown to be precisely these superfields.

Let us  turn now to possible generalizations of our results to
other superparticle models. Increasing $N$ in $D=4$ does not
represent a major complication for our method. For every $N$ there
exist projection operators \cite{GatGriRoeSie}. These could be
used to construct the quantum algebra which should then be
compared with the one arising from the quantization of higher $N$
superparticles. More complicated appears to be the generalization
to higher dimensions where adequate projectors should be
constructed.

In 1946  Snyder \cite{Har46} noted that by introducing a
fundamental length (call it $a$) in physics the commutator of two
space-time operators might not vanish. He indeed proposed the
formula
\begin{gather}
\con{x,y}=\frac{ia^2}{\hbar}L_z
\end{gather}
This should be compared with Pryce  \cite{Pry}\cite{NewWig},
formula (\ref{ConCoord}) valid for systems with internal angular
momentum. Snyder's idea is now a cornerstone of noncommutative
geometry, the implications of which appears to be very important.

There is now a great deal of models with noncommutative
coordinates that are obtained from different routes. As we have
seen non-commutativity of coordinates is not linked to a breaking
of the Lorentz invariance. In this paper we have concentrated
ourselves in the problem of finding a correct set of quantum
operators that satisfy Casalbuoni-Pryce relations for the
superparticle (\ref{conPrime}-\ref{ConCoord}). But the procedure
used  is  not restricted to this system.  We show that
non-commuting operators arise naturally when a set of commuting
operators are projected onto a subspace.

\section{Acknowledgments} This work was supported
by Did-Usb grants Gid-30 and Gid-11 and by Fonacit grant
G-2001000712. N.H thanks S.Okubo for bringing Ref. \cite{Oku} to
his attention.

\setcounter{equation}{0}
\renewcommand{\theequation}{\thesection.\arabic{equation}}
\appendix
\section{Appendix}
In this final section we collect some formulas and conventions that are
necessary in order to follow the calculations. Most conventions are taken
from Wess and Bagger but not all.\\
We set $\sigma^0=\bar\sigma^0=I$ and the rest are like Wess and
Bagger \cite{WesBag}. We always rise and lower indices with the
second index (for example $D^a=\ep^{ab}D_b$). We select
$\ep_{12}=-1$. An important identity is
\begin{gather}
\sigma\ud{\nu}{b\da}\bar\sigma^{\lambda\,\da
a}\sigma\ud{\mu}{a\db}=
-\eta^{\nu\lambda}\sigma\ud{\mu}{b\db}-\eta^{\mu\lambda}\sigma\ud{\nu}{b\db}+\eta^{\mn}
\sigma\ud{\lambda}{b\db}-i\ep^{\nu\lambda\mu\alpha}\sigma_{\alpha\,b\db}\label{BasicTSigma}
\end{gather}
We define
\begin{gather}
{\sigma^\mn}\du{a}{b}=\sigma\ud{\mu}{a\da}\bar\sigma^{\nu\,\da b}-\sigma\ud{\nu}{a\da}\bar\sigma^{\mu\,\da b}\\
\bar\sigma\ud{\mn\,\da}{\db}=\bar\sigma^{\mu\,\da
b}\sigma\ud{\nu}{b\db}-\bar\sigma^{\nu\,\da b}\sigma\ud{\mu}{b\db}
\end{gather}
From where we deduce
\begin{gather}
\sigma^{\mn}\sigma^\alpha+\sigma^\alpha\bar\sigma^{\mn}
=-4i\ep^{\mn\alpha\lambda}\sigma_\lambda\label{Sigmamna}\\
\sigma\ud{\nu}{b\da}\bar\sigma^{\lambda\,\da
a}\sigma\ud{\mu}{a\db}-\sigma\ud{\mu}{b\da}\bar\sigma^{\lambda\,\da
a}\sigma\ud{\nu}{a\db}=-2i\ep^{\nu\lambda\mu\alpha}\sigma_{\alpha\,b\db}\label{AntiBasicSigma}
\end{gather}
Another two important identities
\begin{gather}
\sigma_\beta\bar\sigma_\lambda=-\eta_{\beta\lambda}+\frac{1}{2}\sigma_{\beta\lambda}\label{SimAntiSim}\\
\bar\sigma^{\mu\,\da a}\bar \sigma\du{\mu}{\db b}=-2\ep^{ab}\ep^{\da\db}\label{BarSBarS}
\end{gather}
Covariant derivatives are given by
\begin{gather}
P_\mu=-i\p_\mu\\
\Pi_a=-i\p_a\qquad \bar\Pi_\da=-i\bar\p_\da\\
D_a=\p_a+i\sigma\ud{\mu}{a\db}\bar\Theta^\db\p_\mu=i\Pi_a-P_\mu\sigma\ud{\mu}{a\db}\bar\Theta^\db\label{CovDer1}\\
\bar D_\db=-\bar\p_\db-i\Theta^a\sigma\ud{\mu}{a\db}\p_\mu=-i\bar\Pi_\db+P_\mu\Theta^a\sigma\ud{\mu}{a\db}\label{CovDer2}
\end{gather}
The supersymmetry charges
\begin{gather}
Q_a=\Pi_a-iP_\mu\sigma\ud{\mu}{a\da}\bar\Theta^\da\label{SuperQ1}\\
\bar Q_\da=-\bar\Pi_\da+iP_\mu\sigma\ud{\mu}{a\da}\Theta^a\label{SuperQ2}\ \ .
\end{gather}
Note that we always use a lower case letter for a classical
observable and a capital letter for a quantum operator. In all
cases an operator without any additional label behaves like in the
Schr{\"o}dinger representation. For example
\begin{gather}
X^\mu\Psi=x^\mu\Psi\qquad\Theta^a\Psi=\theta^a\Psi\label{Multi}\\
P_\mu\Psi=-i\p_\mu\Psi\qquad \Pi_a\Psi=-i\p_a\Psi
\end{gather}
Projected operators are often indicated with a label ($A,C$ or
$T$). The momentum operator $P_\mu$ is an exception for this rule
since  we never write $P_\mu^\CC$ for $\Pc P_\mu \Pc$. We think
that this small abuse of notation
cause no confusion.\\
Some useful commutators and anticommutators are \cite{WesBag}
\begin{gather}
\con{\bar D^2,D^a}=-4i\bar \sigma^{\mu\,\db a}\bar D_\db\p_\mu\label{DsquareDa}\\
\con{\bar D^\da,D^2}=-4i\bar \sigma^{\mu\,\da a}D_a\p_\mu\label{DaDsquare}\\
\con{D^2,\bar D^2}=-8iD\sigma^\mu\bar D\p_\mu-16\p^2\label{conD2barD2}\\
\cor{D^2,\bar D^2}=16\p^2+2\bar D_\da D^2\bar D^\da\label{anticonD2barD2}\\
\con{D^2,X^\mu}=2i D^a\sigma\ud{\mu}{a\da}\bar\Theta\label{DsqareX}\\
\con{\bar D^2,X^\mu}=-2i\Theta^a\sigma\ud{\mu}{a\da}\bar D^\da\label{BarDsqareX}
\end{gather}
From theses we may deduce
\begin{gather}
\con{\Pt,\Theta^a}=\frac{1}{8P^2}\cor{D^a,\bar D^2}\label{Pttheta}\\
\con{\Pc,\Theta^a}=\frac{-1}{8P^2}\bar D^2D^a\qquad \con{\Pc,\bar \Theta^\da}=\frac{-1}{8P^2}\bar D^\da D^2\label{Pctheta}\\
\con{\Pa,\Theta^a}=\frac{-1}{8P^2} D^a\bar D^2\qquad \con{\Pa,\bar \Theta^\da}=\frac{-1}{8P^2} D^2\bar D^\da\label{Patheta}\\
\con{\bar D^2,\Pi_b}=2iP_\alpha\sigma\ud{\alpha}{b\db}\bar D^\db\qquad\con{D^2,\bar\Pi_\db}=2iP_\alpha\sigma\ud{\alpha}{b\db}D^b\label{d2Pi}
\end{gather}
An important formula which can be found in \cite{Oku}:
\begin{gather}
\Pt\left(\bar D\bar\sigma^\mu D\right)\Pt P_\mu=-2P^2\Pt\label{OkuDD}
\end{gather}
Note that $0=\con{X^\mu,1}=\con{X^\mu,P^2/P^2}$ implies
\begin{gather}
\con{X^\mu,\frac{1}{P^2}}=-\frac{2i}{(P^2)^2}P^\nu\label{xOverP}
\end{gather}
The covariant derivatives act as interwinding operators
\begin{gather}
\Pa D_b=D_b\Pt\qquad \Pt\bar D_\da=\bar D_\da \Pa\\
\Pt D_a=D_a\Pc\qquad \bar D_\da\Pt=\Pc\bar D_\da\label{interwinding}
\end{gather}

\end{document}